      [1999/12/01 v1.4c Il Nuovo Cimento]
\documentclass{cimento}
\usepackage{amsmath} 
\usepackage{amsfonts} 
\usepackage{amssymb} 

\title{E.C.G. Stueckelberg: a forerunner of modern physics}
\author{F. Cianfrani\from{ins:x}\thanks 
{francesco.cianfrani@icra.it} and O.M. Lecian\from{ins:x}\thanks 
{lecian@icra.it}} 
\instlist{\inst{ins:x} ICRA-International Center for Relativistic Astrophysics\\
Dipartimento di Fisica (G9), Universit\`a  di Roma, ``La Sapienza'', Piazzale Aldo Moro 5, 00185 Rome, Italy.}
\PACSes{{01.30.-y}\hspace{0.3cm}{01.60.+q}\hspace{0.3cm}{01.65.+g} }
\begin{document}

\maketitle

\begin{abstract}
\normalsize
The pioneering work of E.C.G. Stueckelberg is briefly analyzed: the formalism of the Stueckelberg field, able to describe a massive vector field, is reviewed, and some applications are presented. In particular, starting from his very first application, devoted to describe nuclear phenomena known in the Thirties, later attempts to demonstrate the renormalizability of the model are considered. Finally, also string theory and LQG are illustrated to be a suitable scenario for the Stueckelberg field.
\end{abstract}
\section{Biographical notes}
Ernst Carl Gerlach Stueckelberg was born in Basel on February 1st, 1905. In 1926, he got his Ph.D. at Munich under the supervision of Arnold Sommerfeld; then, he qualified as a university lecturer at the University of Zurich, till he became Assistant Professor at Princeton University.\\ 
In 1934 he provided the first covariant perturbation theory for quantum fields. To quote a paper of Lacki et al.\cite{Lacki}, 
\begin{quote}
The approach proposed by Stueckelberg was far more powerful, but was not adopted by others at the time. 
\end{quote}
Then in 1935, before Yukawa\cite{yuk} and by a rather different approach, he proposed to explain nuclear interactions as due to the exchange of vector bosons.\\
Moreover, the evolution parameter theory he presented in 1941 and 1942 is the basis for recent work in Relativistic dynamics.  But his great achievement in 1942 was the interpretation of the positron as a negative energy electron traveling backward in time. Feynmamnn quoted this result in one of his classic papers \cite{fey}.\\
Stueckelberg died in 1984 in Basel.
\section{The Stueckelberg field}
Stueckelberg\cite{Stue,Stue2} developed the only up to known formulation of a renormalizable theory for a massive Abelian boson (for a recent review see\cite{Stuerec}).\\
The first model for massive vector particles was the Proca one\cite{Pro} , which simply produces the extension of the electro-dynamics by the introduction of a mass term; in fact, the Proca Lagrangian density reads 
\begin{equation}
{\mathcal L}_{Proca}=-\frac{1}{2}F^{\dag}_{\mu\nu}F^{\mu\nu}+m^{2}V^{\dag}_{\mu}V^{\mu}.
\end{equation}
It is clear that such a term provides a violation of the Abelian gauge symmetry, and, after several decades, Stueckelberg's work has been recognized as responsible of the renormalizability. After the canonical quantization, one obtains the commutation relations
\begin{equation}
[V_\mu(x);V_\nu(y)]=[V^\dag_\mu(x);V^\dag_\nu(y)]=0;\qquad[V_\mu(x);V^\dag_{\nu}(y)]=-i\bigg(\eta_{\mu\nu}+\frac{1}{m^2}\partial_{\mu}\partial_{\nu}\bigg)\Delta_{m}(x-y)\label{com}
\end{equation}
where the function $\Delta_{m}$ satisfies
\begin{equation}
(\partial^{2}+m^2)\Delta_{m}(x-y)=0
\end{equation}
After 1945, it became clear that the term $\frac{1}{m^2}\partial_{\mu}\partial_{\nu}$ in the commutation relation (\ref{com}) gives rise to ultra-violet divergences, which cannot be eliminated even by the renormalization procedure. However, before the development of the renormalization theory, Stueckelberg\cite{Stue} provided a divergence-free formulation. In his model, the starting point is a Fermi-like Lagrangian density for a complex vector field $A_\mu$, i.e.
\begin{equation}
{\mathcal L}_{A}=-\partial_{\mu}A^\dag_{\nu}\partial^{\mu}A^{\nu}+m^{2}A^\dag_{\mu}A^{\mu};
\end{equation}
since the Hamiltonian density
\begin{equation}
{\mathcal H}_{A}=-\partial_{\mu}A^\dag_{\nu}\partial_{\mu}A^{\nu}-m^{2}A^\dag_{\mu}A^{\mu}
\end{equation}
is not positive definite, one has to impose the analogue of the Gupta-Bleuler condition in electro-dynamics, i.e. that the expectation value on physical states of $\partial^\mu A_\mu$ vanish  
\begin{equation}
<phys'|\partial^\mu A_{\mu}|phys>=0\label{physst} 
\end{equation}
(a sufficient condition is $\partial^\mu A_\mu^{(-)}|phys>=0$, being $A_\mu^{(-)}$ just the positive frequency part). In the Proca case, this condition comes directly from equations of motion.\\
Unlike QED, relation (\ref{physst}) cannot stand, since, from canonical commutation relations, which read
\begin{equation}
[A_\mu(x);A_\nu(y)]=[A^\dag_\mu(x);A^\dag_\nu(y)]=0\qquad[A_\mu(x);A^\dag_\nu(y)]=-i\eta_{\mu\nu}\Delta_{m}(x-y),\label{commA}
\end{equation}
one obtains
\begin{equation}
[\partial^\mu A_\mu(x);\partial^{\nu}A^\dag_{\nu}(y)]=i\partial^2\Delta_{m}(x-y)=im\Delta_{m}(x-y)\neq0.
\end{equation}
Stueckelberg solved this puzzle by introducing a scalar field B(x), whose Lagrangian density reads
\begin{equation}
{\mathcal L}_B=\partial_\mu B^\dag\partial^\mu B-m^2B^\dag B,
\end{equation}
with canonical commutation relations
\begin{equation}
[B(x);B(y)]=[B^\dag(x);B^\dag(y)]=0 \qquad[B(x);B^\dag(y)]=i\Delta_{m}(x-y).\label{commB}
\end{equation}
Hence, the consistency condition on physical states, such that the Hamiltonian density is positive definite, reads as
\begin{equation}
S(x)|phys>=(\partial_\mu A^\mu(x)+mB(x))^{(-)}|phys>=0
\end{equation}
and one can easily demonstrate no contradiction exists with the commutation relations (\ref{commA}), (\ref{commB}). Therefore, the full Stueckelberg Lagrangian density is
\begin{equation}
\mathcal{L}_{Stueck}=-\partial_\mu A_\nu^\dag\partial^\mu A^\nu+m^2 A^\dag_\mu A^\mu+\partial_\mu B^\dag\partial^\mu B-m^2B^\dag B\label{stueclagr}
\end{equation}
which can be cast in the form 
\begin{equation}
\mathcal{L}_{Stueck}=\mathcal{L}_{Proca}(W^\mu)-(\partial_\mu A^{\dag\mu}+mB^{\dag})(\partial_\mu A^\mu+mB);
\end{equation}
being $W^\mu=A^\mu-\frac{1}{m}\partial_\mu B$, it coincides with the Proca Lagrangian density on physical states. However, there is a main difference between the two formulations: while the mass term in $\mathcal{L}_{Proca}$ destroys the gauge symmetry, that in $\mathcal{L}_{Stueck}$ is invariant under Pauli transformations, i.e.
\begin{equation}
\left\{\begin{array}{c} A_\mu\rightarrow A_\mu+\partial_\mu\Lambda\\ B\rightarrow B+m\Lambda \end{array}\right.\qquad (\partial^2+m^2)\Lambda=0.
\end{equation}
A kind of invariance is expected to compensate the introduction of the additional field $B$ and to lower the number of local degrees of freedom to three. In a physical point of view, we can think of the field B(x) as eliminating the scalar term $\partial_\mu A^\mu$ of the vector field. 

\section{The Stueckelberg field and the carriers of nuclear interactions}

The aim of the paper by E.C.G. Stueckelberg was to describe electromagnetic and ``nuclear'' forces (what we would call electromagnetic, weak and strong interactions) within a generalization of the formalism developed for charged particles \cite{Stue0}. This formalism deals with the scalar massive field $A$, which obeys, in presence of matter, the field equation
\small
\begin{equation}
(\partial_{\mu}\partial^{\mu}-l^{2})A=-4\pi J,
\end{equation}
\normalsize
and will be shown to be equivalent to the retarded-potential method, but will offer the advantages of approaching the problem form ``gauge'' point of view ahead of its time\footnote{Throughout this section,we will maintain the original notation adopted by Stueckelberg, in order to appreciate the development of his pioneering intuitions. In particular, the introduction of the Stueckelberg field will be understood from a historical point of viewed, i.e. via the Dirac-Fock-Podolski approximation, rather than from a modern perspective, as reviewed in the previous section.}. The Lagrangian density $\mathcal{L}$ reads
\small
\begin{equation}
\mathcal{L}=-\frac{1}{8\pi}\left[\left( \frac{\partial A}{\partial x},\frac{\partial A^{*}}{\partial x}\right)+l^{2}A^{*}A\right]+\frac{1}{2}\left[A^{*}J+\left(\frac{\partial A^{*}}{\partial x},S\right)+c.c.\right],
\end{equation}
\normalsize
where $A$ and $A^{*}$ are treated like independent quantities. The ``effective'' current $J_{eff}$ rewrites $J_{eff}=J-\left(\frac{\partial}{\partial x},s\right)$, as a function of the polarization vector $S$. From the conjugate momentum $P=\frac{\partial \mathcal{L}}{\partial \dot{A}}$, the Hamiltonian density $H$ is found, and, for later purposes, it will be expressed as
\small
\begin{equation}
H=\int d^{3}x\left( -\mathcal{L}+\dot{A}P+\dot{A}^{*}P^{*}\right)\equiv\int d^{3}x\left( \mathcal{W}+\mathcal{V}\right)\equiv W+V,
\end{equation}
\normalsize
where $\mathcal{W}\equiv \frac{1}{8\pi}\left[\left( \frac{\partial A}{\partial x},\frac{\partial A^{*}}{\partial x}\right)+l^{2}A^{*}A\right]+8\pi c^{2}P^{*}P$, and $\mathcal{V}\equiv -\frac{1}{2}\left[A^{*}J+\left(\frac{\partial A^{*}}{\partial x},S\right)+c.c.\right]-4\pi c\left( PS_{0}+cc \right)$,respectively. Motion equations follow from the introduction of the operator $K$, that allows one to get a straightforward definition of $J$ and $S$:
\small
\begin{equation}
\left(\frac{i}{h}\right)\left[K,P^{*}\right]=-\frac{\delta K}{\delta A^{*}}=\frac{1}{2}\left[J-\left( \frac{\partial}{\partial x},S\right)\right]
\end{equation}
\begin{equation}
\left(\frac{i}{h}\right)\left[K,A\right]=-\frac{\delta K}{\delta P}=-4\pi cS_{0}.
\end{equation}
\normalsize 
$J$ and $S$ are functions of the canonical variables $p$ and $q$, which describe the matter distribution, and obey the (classical) equations of motion $\dot{p}=(i/h)[K,p]$ and $\dot{q}=(i/h)[K,q]$.
The quantum theory can be implemented  by solving the Schroedinger equation $H\Psi(t)=ih\frac{\partial\Psi(t)}{\partial t}$. To this end, the functional $\Psi'(T,t)$ for the wave function is introduced, such that $\Psi'(t,t)\equiv\Psi(t)$, and, accordingly, the functional $K(T,t)$, such that $K(t,t)\equiv K$ ($K$ does not depend on $t$ explicitly). Since the functional $\Psi'$ must satisfy simultaneously the two Schroedinger equations
\small 
\begin{equation}\label{W}
C_{T}\Psi'(T,t)=\left(W+\frac{h}{i}\frac{\partial}{\partial T}\right)\Psi'(T,t)=0,
\end{equation}
\begin{equation}\label{K}
C_{t}\Psi'(T,t)=\left(K'(T,t)+\frac{h}{i}\frac{\partial}{\partial T}\right)\Psi'(T,t)=0,
\end{equation}
\normalsize
the wave function $\Psi$ is defined by $H=W+K$. The request that the two Schroedinger equations be simultaneously solvable leads to the vanishing commutation relation between the operators defined in (\ref{W}) and (\ref{K}), i.e. $\left[C_{T},C_{t}\right]=0$, from which the expression for K
\small
\begin{equation}
K'(T,t)=e^{iW(t-T)/h}Ke^{-iW(t-T)/h} 
\end{equation}
\normalsize
is found; consequently, $\Psi$ admits the formal solution
\small
\begin{equation}
\Psi(T,t)=e^{-iWT/h}\psi(t),
\end{equation}
\normalsize
where $\psi(t)$ satisfies the Schroedinger equation
\small
\begin{equation}\label{313}
K''(T)\psi(t)=ih \frac{\partial \psi(t)}{\partial t},
\end{equation}
\normalsize 
with $K''(t)=e^{iWt/h}Ke^{-iWt/h}$. If matter distribution is described in the configuration space, with coordinates $\left\{q^{s}\right\}$, rather than by means of matter fields, the Hamiltonian operator $K$ can be rewritten as the sum of two terms,
\small
\begin{equation}\label{sum}
K=\sum_{s}K_{s}\equiv\sum_{s}\left(R_{s}+V_{s}\right),
\end{equation}
\normalsize
where the former depends on the $\left\{q^{s}\right\}$'s only, $R_{s}\equiv R_{s}(q^{s})$, while the latter is a function of both the coordinates $\left\{q^{s}\right\}$ and the field $A$, $V_{s}\equiv V_{s}(q^{s}, A((\vec{x}),t))\equiv V_{s}(t)$, as it will be explained in the following. Eq. (\ref{313}) now reads
\begin{equation}
\left(K''+\frac{h}{i}\frac{\partial}{\partial t}\right)\psi=\left(R+V+\frac{h}{i}\frac{\partial}{\partial t}\right)\psi=0,
\end{equation}
where the sum (\ref{sum}) is taken into account. This Schroedinger equation contains the ``current term'' $V_{s}$, which contains, on its turn, the field $A$: assuming that this term is proportional to a small number, a series expansion will be performed in order to obtain the approximated expression for $\psi$, i.e. $\psi=\psi^{0}+\psi^{1}+\psi^{2}+...$:
\small
\begin{equation}\label{43}
\left(R+\frac{h}{i}\frac{\partial}{\partial t}\right)\psi^{0}=0,
\end{equation}
\begin{equation}
\left(R+\frac{h}{i}\frac{\partial}{\partial t}\right)\psi^{1}+V\psi^{o}=0,
\end{equation}
\normalsize
and so on. Collecting the terms for the proper approximation order, one finds
\small
\begin{equation}
V_{s}(t)\psi^{1}=\sum_{r}U_{sr}\psi^{0},
\end{equation}
\normalsize
so that
\small
\begin{equation}
\left(R+\frac{h}{i}\frac{\partial}{\partial t}\right)(\psi^{1}+\psi^{2})+\left(V+\sum_{s}\sum_{r}U^{rs}\right)\psi^{0}=0,
\end{equation}
\normalsize
%\begin{equation}
%\left(R+V+\sum_{s}\sum_{r}U^{rs}+\frac{h}{i}\frac{\partial}{\partial t}\right)\psi=0
%\end{equation}
where the term $U_{sr}+U_{rs}$ is recognized as the first order approximation of the ``exchange energy''. Since the wave function $\psi$ must describe the distribution of all the particles, a ``multi-time functional'' $\psi(t^{1}, ..., t^{s}, ..., t^{n})$ can be defined, such that, as previously, $\psi(t, ...,t,..., t)\equiv\psi(t)$, so that the Schroedinger equation for the wave functional reads
\begin{equation}\label{48}
\left(R_{s}+V_{s}(t^{s})+\frac{h}{i}\frac{\partial}{\partial t_{s}}\right)\psi(t^{1}, ..., t^{s}, ..., t^{n})=0.
\end{equation}
A reference frame can be found, where $R_{s}$ does not depend on time explicitly; here, the eigenvalue equation 
\begin{equation}
f(R_{s})=u_{\nu_{1}...\nu_{n}}f(h_{\nu_{s}})
\end{equation}
holds, $u_{\nu_{s}}$ being time-independent functions; the corresponding time-dependent functions are $v_{\nu_{1}...\nu_{n}}=u_{\nu_{1}...\nu_{n}}e^{-i\sum_{s}\nu_{s}t^{s}}$, which satisfy (\ref{48}) at the $0^{th}$ order. The functions $w$ are defined as functionals of $t^{s}$, $q^{s}$ and the fields $A$, and their time dependence is given by
\small
\begin{equation}
w=\sum_{\omega_{1}}...\sum_{\omega_{n}}e^{-i\sum_{s}\omega_{s}t^{s}}w_{\omega_{1}...\omega_{n}},
\end{equation}
\normalsize
so that
\small
\begin{equation}\label{omega}
f\left(R_{s}+\frac{h}{i}\frac{\partial}{\partial t^{s}}\right)w_{t^{1}...t^{n}}=\sum_{\omega_{1}}...\sum_{\omega_{n}}e^{-\sum_{s}\omega_{s}t^{s}f(R_{s}-\omega^{s}}w_{\omega_{1}...\omega_{n}}.
\end{equation}
\normalsize
It is now possible to solve the system (\ref{43}), so that, at the $1^{st}$ order, the functional $\psi$ reads
\small
\begin{equation}
\psi^{1}=-\sum\left(R_{r}+\frac{h}{i}\frac{\partial}{\partial t^{r}}\right)^{-1}V_{r}(t^{r})\psi^{0},
\end{equation}
\normalsize
where $\psi^{0}(t^{1}...t^{n})=e^{-i\sum_{r}R_{r}(t^{r}-t)}\psi^{0}(t)$ : it is easy to verify that $\psi^{0}$ must be a linear combination of the eigenfunctions $u_{\nu_{1}...\nu_{n}}$, with time-dependent coefficients $e^{-i\sum\nu_{r}t}$. One is therefore interested only in the $1^{st}$ order time-independent matrix elements of the operator
\begin{equation}
U_{sr}=-\left[\left(R_{r}+\frac{h}{i}\frac{\partial}{\partial t^{r}}\right)^{-1}V_{s}(t^{s})V_{r}(t^{r})e^{-i\sum_{m}R_{m}(t^{m}-t)}\right]_{t^{1}=t^{2}=...=t}, 
\end{equation}
which are found by considering the integral
\small
\begin{equation}
U^{sr}_{\nu'\nu}=\int dq^{1}...\int dq^{n}\int dAv^{*}_{\nu_{1}'...\nu_{n}'}\left(R_{r}+\frac{h}{i}\frac{\partial}{\partial t^{r}}\right)^{-1}V_{s}(t^{s})V_{r}(t^{r})v_{\nu_{1}'...\nu_{n}'}:
\end{equation}
\normalsize
because of (\ref{omega}), the condition $(\nu_{s}'-\omega_{s})+(\nu_{r}'-\omega_{r})+\sum_{m\neq s,r}(\nu_{m}'-\omega_{m})=0$ must be fulfilled. After standard manipulation, one finds that the time-independent matrix elements are given by the operator
\begin{equation}
\int_{t}^{\infty}dt^{r}e^{iR_{r}(t^{r}-t)}[V_{s}(t),V_{r}(t^{R}]e^{-iR_{r}(t^{r}-t)},
\end{equation}
where
\small
\begin{equation}
V_{s}=-\frac{1}{2}\int d^{3}x\left(A^{*}(x),J_{s}(x)\right)+\left(\frac{\partial A^{*}(x)}{\partial x},S_{s}(x) \right)+c.c.+\mathcal{O}(A^{2}):
\end{equation}
\normalsize
the terms in $A^{2}$ must be neglected, since, at this order, no quantity has been developed up to higher powers of the field. This way, the quantities $J_{s}$ and $S_{s}$ do not depend on the fields any more,  and commute with them. In particular, one finds for the current $J_{s}$ the formal solution
\small
\begin{equation}
J_{s}(y)=e^{iR_{r}\frac{y_{0}-ct}{ch}}J_{s}(\vec{y})e^{-iR_{r}\frac{y_{0}-ct}{ch}},
\end{equation}
\normalsize
so that the interaction operator reads
\small
\begin{equation}
U_{rs}+U_{sr}=-\frac{1}{2}\int d^{3}y \left[J_{s}(\vec{x})A_{r}(x)^{*}+\left(S_{s}(\vec{x}), \frac{\partial A_{r}(x)^{*}}{\partial x}\right)\right]_{x_{0}=ct},
\end{equation}
\normalsize
where
\small
\begin{equation}
A_{r}(x)=\int_{x_{0}}^{\pm\infty}dy_{0}\int d^{3}x\left[J_{r}(y)D(x-y)+\left(S_{r}(y), \frac{\partial D(x-y)}{\partial y}\right)\right]=\int_{x_{0}}^{\pm\infty}dy_{0}\int d^{3}y J_{r}^{eff}(y)D(x-y),
\end{equation}
\normalsize
with $J^{eff}$ defined as previously.\\
The generalization of this formalism to a ``many-component'' field \cite{Stue} (what we would call a vector field) can be accomplished via the substitution of the field $A$ with the field $A_{i}$, $i=0,1,2,3$, so that, for example, the scalar product $A^{*}A$ is replaced by $\sum_{i}\epsilon_{i}A^{*}_{i}A_{i}$, where $\epsilon_{0}\equiv -1$, $\epsilon_{1,2,3}\equiv 1$, and so on. New commutation relations have to be introduced, such as
\small
\begin{equation}
\left[A_{i}^{*}(x),A_{j}(y)\right]=2ihc\epsilon_{i}\delta_{ij}D(x-y).
\end{equation}
\normalsize
In order to have a positive-definite energy density for the field $A_{i}$, the new field $B$, the so-called Stueckelberg field, has to be introduced; in fact, the energy density $\mathcal{H}=\sum_{i}\epsilon_{i}\mathcal{H}(A_{i})$, where
\small
\begin{equation}\label{82}
\mathcal{H}(A)=\frac{1}{8\pi}\left(\sum_{k}\frac{\partial A^{*}}{\partial x_{k}}\frac{\partial A}{\partial x_{k}}+l^{2}A^{*}A\right)
\end{equation}
\normalsize
contains a negative term when $i=0$. The way followed by Stueckelberg in order to determine this term is the ``Dirac-Fock-Podolski approximation'' $\left(\frac{\partial}{\partial x},A\right)\psi=0$, which can be interpreted as a Gupta-Bleuer condition on the divergence of $A$: when a mass term is introduced, and when a vector field is taken into account, the approximation reads
\small
\begin{equation}
-\frac{\partial A^{*}_{0}}{\partial x_{0}}\frac{\partial A_{0}}{\partial x_{0}}\psi=\left(-div \vec{A}^{*} div \vec{A}-l(B^{*}div\vec{A}+div\vec{A}^{*}B)-l^{2}B^{*}B\right)\psi
\end{equation}
\normalsize
that eliminates the negative term in (\ref{82}). After standard manipulation the energy density for the two fields, $\mathcal{H}'(A,B)$ reads

\vspace{0.2cm}

$8\pi\mathcal{H}'=\left(rot \vec{A}^{*},rot\vec{A}\right)+\left(grad A_{0}^{*}+\frac{\partial \vec{A}^{*}}{\partial x_{0}},grad A_{0}+\frac{\partial \vec{A}}{\partial x_{0}}\right)+$

\vspace{0.2cm}

$\left(lA_{0}^{*}-\frac{\partial B^{*}}{\partial x_{0}}\right)\left(lA_{0}-\frac{\partial B}{\partial x_{0}}\right)+\left(l\vec{A}^{*}+grad B^{*}\right)\left(l\vec{A}+grad B\right):$

\vspace{0.2cm}

if the new potential $\mathbf{\phi}_{i}=A_{i}+\epsilon_{i}l^{-1}\frac{\partial B}{\partial x_{i}}$ is introduced, the energy desity rewrites
\small
\begin{equation}
\mathcal{H}'=\frac{1}{8\pi}\left[\left(\vec{F}^{*},\vec{F}\right)+\left(\vec{G}^{*},\vec{G}\right)+l^{2}\left(\vec{\mathbf{\phi}}^{*},\vec{\mathbf{\phi}}\right)+l^{2}\mathbf{\phi}_{0}^{*}\mathbf{\phi}_{0}\right],
\end{equation}
\normalsize
where $\vec{F}\equiv\left\{F_{01},F_{02}F_{03}\right\}$ and $\vec{G}\equiv\left\{F_{23},F_{31}F_{12}\right\}$, $F_{ij}$ being the field strength, $F_{ij}\equiv \epsilon_{i}\frac{\partial \mathbf{\phi}_{j}}{\partial x_{i}}- \epsilon_{j}\frac{\partial \mathbf{\phi}_{i}}{\partial x_{j}}\equiv\epsilon_{i}\frac{\partial A_{k}}{\partial x_{i}}- \epsilon_{j}\frac{\partial A_{i}}{\partial x_{j}}$.\\
If the same calculation as the case of the scalar field is followed, motion equations for spinors and bosons are obtained. E.C.G. Stueckelberg, in fact, wanted set up a unifying theory for scattering and decay processes, within the framework of boson ``gauge'' fields: he achieved this task by taking into account the then-known particles and interactions, by hypothesizing generalized-``charge'' conservation, and by predicting, from his calculation, the existence of new particles and information about their masses \cite{Stue2}. Unfortunately, not all leptons had already been observed yet, and, consequently, the notion of leptonic and barionic number, as well as the distinction of weak and strong interactions, had not already been introduced at that time, but he laid  the theoretical foundation of gauge theories. According to the results of the experiments, he classified the known ``spinor'' particles according to their scattering and decay properties  by attributing them electric and ``heavy'' charges, so that electrons, neutrinos, protons and neutrons are referred to as $e(1,0)$, $n(0,0)$, $P(1,1)$ and $N(0,1)$, respectively. As in modern gauge theories, interaction between these charges are described by boson fields, which follow directly from the eigen-value equations for the generators, so that he predicts four such fields, that, according to the interaction they carry, are classified as \textbf{e} $(1,0)$,  \textbf{n}$(0,0)$, \textbf{P} $(1,1)$ and \textbf{N} $(0,1)$, respectively. Therefore, the processes mediated by these fields are\\
\begin{itemize}
{\item 1) processes mediated by \textbf{n}$(0,0)$:\\
the only processes described by these fields are of the type
\small
\begin{equation}
S\rightarrow S' + \textbf{n}(0,0),
\end{equation}
\normalsize
where $S$ can be referred to any kind of spinor. In Stueckelberg's interpretation, a better understanding of the proton-proton and neutron-neutron interactions in atomic nuclei could be achieved by means of the real field \textbf{n}$(0,0)$.}
{\item 2) processes mediated by the field \textbf{e}$(1,0)$:\\
these processes are $\beta$-decays:
\small
\begin{equation}
P(1,1)\leftrightarrow N(0,1)+\textbf{e}(1,0),
\end{equation}
\normalsize
together with
\small
\begin{equation}
e(1,0)\leftrightarrow (-n(0,0))+\textbf{e}(1,0)
\end{equation}
\normalsize
describe a nuclear decay, where the notion of anti-particle follows from the mathematics of the model.} 
{\item 3) processes mediated by \textbf{N} $(0,1)$:\\
the reactions
\small
\begin{equation}\nonumber
N(0,1)\leftrightarrow (-n(0,0))+\textbf{N}(0,1),
\end{equation}
\begin{equation}
P(1,1)\leftrightarrow e(1,0)+\textbf{N}(0,1)
\end{equation}
\normalsize
lead to estimate the mass of the particle \textbf{N} $(0,1)$: since the proton is a stable particle, the mass of \textbf{N} $(0,1)$ must be greater than the difference of the masses of the proton and of the electron; furthermore, because of statistics, the mass of the particle \textbf{N} $(0,1)$ must be greater than the neutron mass, and it must be an instable particle, whose decay mode is
\small
\begin{equation}
\textbf{N}(0,1)\rightarrow P(1,1)+(-e(1,0)).
\end{equation}
\normalsize
}
{\item 4) processes mediated by \textbf{P} $(1,1)$:\\
\small
\begin{equation}
P(1,1)\leftrightarrow (-n(0,0))+\textbf{P}(1,1),
\end{equation}
\normalsize}
\end{itemize}
so that the mass of the particle \textbf{P}$(0,1)$ must be greater than the proton mass.\\
A modern approach to the Electroweak model via the Stueckelberg field is proposed in \cite{Stuerec}.
\section{The Stueckelberg field beyond Stueckelberg}
Application of this formalism was at first devoted to demonstrate its renormalizability. In this sense, Zimmermann\cite{Zim} started to study Stueckelberg Lagrangian (\ref{stueclagr}), and its invariance under the Pauli gauge transformations. At the end, the Stueckelberg massive Abelian model was proved to be renormalizable and unitary by Lowenstein and Schroer in 1972\cite{LB} . We want to stress that this implies that the Stueckelberg model is the only way to give a mass to an Abelian boson, without a spontaneous symmetry breaking mechanism. Therefore, there were several attempts to apply the theory to the non-Abelian case, in order to to furnish an alternative to the Higgs boson in the Standard Model.\\
In 1988, Delbourgo, Twisk and Thompson\cite{DTT} first proved that the original Stueckelberg theory for neutral massive vector fields is invariant under nilpotent BRST transformations, which ensures unitarity and renormalizability. Their work clearly illustrated that the key point, to avoid divergences, is the invariance under Pauli transformations. Then, they also analyzed the extension to non-Abelian fields. They noticed that renormalizability and unitarity seem to be competing qualities of massive non Abelian theories, so they argued:
\begin{quote}
``Finally, it must be admitted that the Higgs mechanism remains the most complete method for giving mass to the vector bosons''.
\end{quote}
But extension of the Standard Model, such to contain a Stueckelberg field, are again under investigation\cite{StuSM} .\\ 
However, Stueckelberg theory for massive bosons found application also very far from its natural context.\\ 
An example is given by the work of Ramond\cite{Ram} , who applied the scheme to obtain the fully covariant and gauge invariant field theory for free open bosonic strings in 26 dimensions. In fact, Stueckelberg fields naturally arise and are shown to be unrestricted for the most general gauge transformations.\\ 
To quote his own words:
\begin{quote}
``It should be clear that Stueckelberg field leads to much simpler looking expressions''.\\
\end{quote}                       
Moreover, also very recent attempts to introduce a massive Abelian field in Loop Quantum Gravity deal with the Stueckelberg field.\\
Hence, Helesfai\cite{Hel} stressed how, in such a context, the application of the Stueckelberg formalism is very useful since no second class constraint arise and the Hamiltonian is a linear combination of constraints (after quantization, the Proca field leads to a Hamiltonian that is quadratic in the Lagrange multipliers).
In fact, the Hamiltonian reads
\begin{equation}
H=\int_{\Sigma}(N\mathcal{H}+N^{a}\mathcal{H}_{a}+A^b_0G_b+A_0\underline{G})d^3x
\end{equation}
being

\vspace{0.2cm}

$\mathcal{H}=\frac{1}{\sqrt{q}}tr(2[K_a;K_b]-F_{ab})[E^a;E^b]+\frac{q_{ab}}{2\sqrt{q}}(\underline{E}^{a}\underline{E}^{b}+\underline{B}^{a}\underline{B}^{b})+\frac{\pi^2}{2\sqrt{q}m^2}+\frac{\sqrt{q}m^2}{2}q^{ab}(\underline{A}_a+\partial_a\phi)(\underline{A}_b+\partial_b\phi)$

\vspace{0.2cm}

$\mathcal{H}_{a}=F^{j}_{ab}E^b_j+\epsilon_{abc}\underline{E}^b\underline{B}^c+(\underline{A}_a+\partial_a)\pi$

\vspace{0.2cm}

$\underline{G}=D_a\underline{E}^a-\pi$

\vspace{0.2cm}

$G_b=D_a E^a_b$

\vspace{0.2cm}

The quantization is performed on the Hilbert space
\begin{equation}
\mathcal{H}=L_2(\bar{\mathcal{A}}_{SU(2)},d\mu_{SU(2)})\otimes L_2(\bar{\mathcal{A}}_{U(1)},d\mu_{U(1)})\otimes
L_2(\bar{\mathcal{U}}_{U(1)},d\mu_{U(1)})
\end{equation}
for which a basis is given by the generalized spin network functions
\begin{eqnarray}
|S>_{\gamma,\vec{j},\vec{\rho},\vec{l},\vec{m}}=|T(A)>_{\gamma,\vec{j},\vec{\rho}}\otimes |F(\underline{A})>_{\gamma,\vec{l}}\otimes 
|D(U)>_{\gamma,\vec{m}}.
\end{eqnarray}
In this context, the mass $m$ is a coupling constant and is very similar to the Immirzi parameter (in the quantum regime, it enters the Hamiltonian in a non-trivial way).

\section{Brief concluding remarks}
Among the brilliant results accomplished by Stueckelberg, the formulation of a divergence-free model for massive vector fields has been one of the most prolific ideas in modern Physics. In fact, despite the Proca formulation, his intuition of the need to maintain a gauge invariance in the theory has been the key to the later-recognized renormalizability. Moreover, the modernity of his approach  relies on the preference of a gauge symmetry rather than phenomenological speculations, such as the Yukawa formulation \cite{yuk}, and renders the Stueckelberg field a suitable tool also in current achievements of theoretical Physics, i.e. String theory and LQG.     

\section{Acknowledgment}
We wish to thank Prof. Remo Ruffini and Dr. Giovanni Montani for having attracted our attention to the pioneering character of Stueckelberg's work.

\end{document}